\documentclass[useAMS,usenatbib]{mn2e}

\usepackage{url} % for \url{}
\usepackage{subfig}
\usepackage{graphicx}

\bibpunct{(}{)}{,}{a}{}{,} %%% !!! This is to remove coma between Author and year !!!

\title[The core shift effect in the blazar 3C\,454.3]{%
The core shift effect in the blazar 3C\,454.3}
\author[]{\parbox{\textwidth}{A.~M.~Kutkin$^{1}$\thanks{E-mail:
kutkin@asc.rssi.ru},
K.~V.~Sokolovsky$^{1,2}$, M.~M.~Lisakov$^{1}$,
Y.~Y.~Kovalev$^{1,3}$, T.~Savolainen$^{3}$, P.~A.~Voytsik$^{1}$,
A.~P.~Lobanov$^{3}$, H.~D.~Aller$^{4}$, M.~F.~Aller$^{4}$,
A.~Lahteenmaki$^{5,6}$, M.~Tornikoski$^5$,
A.~E.~Volvach$^7$, L.~N.~Volvach$^7$}\vspace{0.4cm}\\
\parbox{\textwidth}{$^{1}$Astro Space Center of Lebedev Physical Institute, Profsoyuznaya Str.
84/32, 117997 Moscow, Russia\\
$^{2}$Sternberg Astronomical Institute, Moscow State University,
Universitetskii~pr. 13, 119992 Moscow, Russia\\%
$^{3}$Max-Planck-Institut f\"ur Radioastronomie, Auf dem H\"ugel 69,
D-53121 Bonn, Germany\\%
$^4$University of Michigan, Astronomy
Department, Ann Arbor, MI 48109-1042 USA\\%
$^5$Aalto University Mets\"{a}hovi Radio Observatory, 114, Kylm\"{a}l\"{a}, 02540, Finland\\%
$^6$Department of Radio Science and Engineering, 13000, FI-00076 AALTO, Finland\\%
$^7$Radio Astronomy Laboratory of the Crimean Astrophysical
Observatory, Katsiveli, Crimea, 98688 Ukraine}\\%
}
\begin{document}

\date{Accepted 2013 XXXXX XX. Received 2013 XXXXX XX; in original form 2013 XXXXX XX}

\pagerange{\pageref{firstpage}--\pageref{lastpage}} \pubyear{2013}

\maketitle

\label{firstpage}

\begin{abstract}
Opacity-driven shifts of the apparent VLBI core position with
frequency (the ``core shift'' effect) probe physical conditions in
the innermost parts of jets in active galactic nuclei. We present
the first detailed investigation of this effect in the brightest
$\gamma$-ray blazar 3C\,454.3 using direct measurements from
simultaneous 4.6--43\,GHz VLBA observations, and a time lag analysis
of 4.8--37\,GHz lightcurves from the UMRAO, CrAO, and Mets\"{a}hovi
observations in 2007--2009. The results support the standard
K{\"o}nigl model of jet physics in the VLBI core region. The
distance of the core from the jet origin $r_\mathrm{c}(\nu)$, the
core size $W(\nu)$, and the lightcurve time lag $\Delta T(\nu)$ all
depend on the observing frequency $\nu$ as $r_\mathrm{c}(\nu)
\propto W(\nu) \propto \Delta T(\nu) \propto \nu^{-1/k}$. The
obtained range of $k=0.6-0.8$ is consistent with the synchrotron
self-absorption being the dominating opacity mechanism in the jet.
The similar frequency dependence of $r_\mathrm{c}(\nu)$ and $W(\nu)$
suggests that the external pressure gradient does not dictate the
jet geometry in the cm-band core region. Assuming equipartition, the
magnetic field strength scales with distance $r$ as $B =
0.4(r/1\,pc)^{-0.8}$\,G. The total kinetic power of
electron/positron jet is about $10^{44}$~ergs~s$^{-1}$.
\end{abstract}

\begin{keywords}
galaxies: active -- galaxies: jets -- radio continuum: galaxies --
quasars: individual: 3C454.3
\end{keywords}

\section{Introduction}

3C\,454.3, also known as PKS\,B2251$+$158
($\alpha_\mathrm{J2000}$=22:53:57.747940
$\delta_\mathrm{J2000}$=$+$16:08:53.56074\footnote{Position from the
  Radio Fundamental Catalog version \texttt{rfc\_2013b}, see:
  \url{http://astrogeo.org/rfc/}}; $z=0.859$
\citealt{1991MNRAS.250..414J}) and nicknamed the ``Crazy Diamond''
by the {\it AGILE} team for its brightness and unpredictable
behavior \citep{2010ApJ...712..405V}, is a prominent member of the
blazar class of active galactic nuclei. Like other blazars,
3C\,454.3 contains a relativistic plasma jet pointed close to our
line of sight. As a result of relativistic beaming, synchrotron
radiation from the jet dominates the blazar's observed energy output
from radio to infrared and optical bands
\citep{1980ApJ...235..386M}. During a low flux density state, when
the jet emission is weak, a prominent ultraviolet excess is observed
which is attributed to thermal emission of the accretion disk
\citep[e.g.,][]{1988ApJ...326L..39S,2009A&A...504L...9V,2011A&A...534A..87R}.
%\citep{1988ApJ...326L..39S,2006A&A...453..817V,2007A&A...473..819R,2009A&A...504L...9V,2011A&A...534A..87R}.
The bright X-ray to GeV emission of 3C\,454.3 is likely due to
inverse Compton scattering of photons from an external source (a
broad-line region gas, accretion disk or dusty torus) by
relativistic leptons in the jet (e.g.,
\citealt{2009ApJ...692...32D}). In 2008--2010 3C\,454.3 showed a
spectacular series of GeV flares becoming the brightest object in
the $\gamma$-ray sky
\citep{2009ApJ...699..817A,2010ApJ...718..455S,2010ApJ...721.1383A,2011MNRAS.410..368B,2011ApJ...733L..26A,2011ApJ...736L..38V}.
The $\gamma$-ray flares were echoed in other bands
\citep{2009A&A...504L...9V,2009ATel.2329....1S,2010ApJ...712..405V,2010ApJ...715..362J,2010ApJ...716L.170P},
but the corresponding optical flares were not exceptional for this
blazar \citep{2010ATel.3047....1K}. So far, the object was not
detected in TeV band \citep{2009A&A...498...83A}.

\begin{figure*}
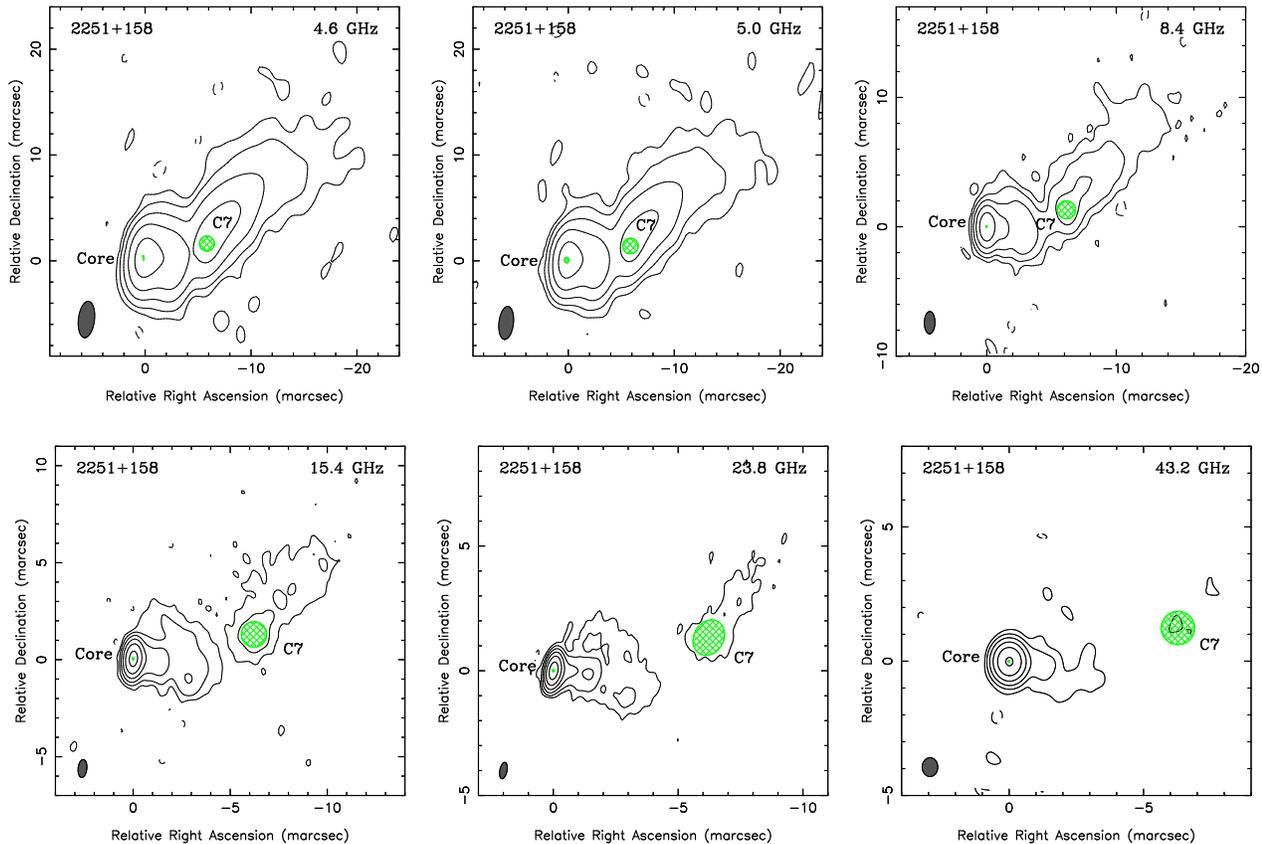

%  \centering
  \includegraphics[width=0.3\textwidth, angle=270, trim=0cm 0cm 0cm 0cm, clip]{2251+158.4.6GHz.2008-10-02.bw.ps}
  ~~
  \includegraphics[width=0.3\textwidth, angle=270, trim=0cm 0cm 0cm 0cm, clip]{2251+158.5.0GHz.2008-10-02.bw.ps}
  ~~
  \includegraphics[width=0.3\textwidth, angle=270, trim=0cm 0cm 0cm 0cm, clip]{2251+158.8.4GHz.2008-10-02.bw.ps}
  \\
  \vskip 0.5cm
  \includegraphics[width=0.3\textwidth, angle=270, trim=0cm 0.1cm 0cm 0cm, clip]{2251+158.15.4GHz.2008-10-02.bw.ps}
  ~~
  \includegraphics[width=0.3\textwidth, angle=270, trim=0cm 0cm 0cm 0cm, clip]{2251+158.23.8GHz.2008-10-02.bw.ps}
  ~~
  \includegraphics[width=0.3\textwidth, angle=270, trim=0cm 0cm 0cm 0cm, clip]{2251+158.43.2GHz.2008-10-02.bw.ps}

  \caption{Naturally weighted total intensity CLEAN images of
    3C\,454.3.  The image peaks are 3.3, 3.4, 6.3, 12.0, 15.8,
    23.5\,Jy/beam, for the 4.6, 5.0, 8.4, 15.4, 23.8, and 43.2\,GHz
    frequencies, respectively.  The contours starting at $6$\,mJy/beam
    are plotted with a factor of 4 steps at all images.  The beam is
    plotted in the lower left corner of each image at the half-power
    level.  The 8.4\,GHz image is almost identical to the 8.1\,GHz
    one, which is not shown here.  The observing
    epoch is 2 October 2008.}
  \label{fig:colormap}
\end{figure*}

% The core
Radio observations with the Very Long Baseline Interferometry (VLBI)
technique provide images of extragalactic jets with spatial
resolution of an order of a parsec (e.g.,
\citealt{1996ASPC..100...97P,1997ARA&A..35..607Z,2006astro.ph.10712Z,2010arXiv1010.2856L}).
The structure of most blazars, including 3C\,454.3
(Fig.~\ref{fig:colormap}), is dominated by a bright, unresolved or
barely-resolved feature called the core. The core has a flat or
inverted radio spectrum, characteristic of optically-thick
synchrotron emission
\citep[e.g.,][]{2006MNRAS.367.1083K,2012MNRAS.423..756P}.
%, and may show flux and polarization variability on timescales
shorter than any other jet structure.  Evidence is growing that
processes close to the VLBI core are responsible for the high-energy
emission of blazars
\citep[e.g.,][]{2009ApJ...696L..17K,2010ApJ...722L...7P,2011A&A...532A.146L,2012A&A...537A..70S,2012ApJ...758...72W}.
The physical nature of the parsec-scale core is still being debated
\citep{2006AIPC..856....1M,2008ASPC..386..437M}, however the most
widely accepted interpretation is that the core is a surface in a
continuous flow where the optical depth of jet's synchrotron
radiation $\tau_\nu \approx 1$ -- ``photosphere''
\citep{1979ApJ...231..299R,1981ApJ...243..700K,1995ApJ...443...35Z}.

% The core shift
The standard jet model
\citep{1979ApJ...232...34B,1981ApJ...243..700K} predicts that the
apparent position of the photosphere (and, therefore, the VLBI core
if the above interpretation is correct) depends on the observing
frequency. This is known as the ``core shift'' effect and it was
first observed by \cite{1984ApJ...276...56M} in the quasar pair
1038$+$528\,A,\,B and later in other sources by, among others,
\cite{1995ApJ...443...35Z,1998A&A...330...79L,2000ivsg.conf..342P,2008A&A...483..759K,2009MNRAS.400...26O,2011A&A...532A..38S,2011Natur.477..185H,2012MNRAS.420..542A},
and \cite{2012A&A...545A.113P}. Measurements of the
frequency-dependent core position shift may provide important
information about the physical conditions and structure of ultra
compact blazar jets
\citep{1998A&A...330...79L,2005ApJ...619...73H,2009MNRAS.400...26O}.
Such observations may constrain the nature of the absorbing
material: synchrotron self-absorption (SSA) within the jet plasma
vs. free-free absorption in thermal plasma surrounding the jet.
While SSA appears to be the dominating opacity mechanism in blazar
jets \citep{2011A&A...532A..38S}, free--free absorption is found in
relativistic jet sources viewed at large angles to the line of
sight: Cyg~A \citep{2008evn..confE.108B} and NGC\,1052
\citep{2004A&A...426..481K}.

The core shift effect has important consequences for ultra-high
precision astrometry
\citep{2005astro.ph..5475R,2009A&A...505L...1P}, radio (VLBI) to
optical (\textit{GAIA}) reference frame alignment
\citep{2008A&A...483..759K} and spacecraft navigation with VLBI. It
should be taken into account when constructing VLBI spectral index
\citep[e.g.,][]{2001ApJ...550..160M,2008A&A...483..759K} and Faraday
rotation maps \citep[e.g.,][]{2012AJ....144..105H}.

This paper presents the first detailed study of the core shift
effect in the quasar 3C\,454.3 using both multi-frequency VLBI
results and single-dish lightcurves. Core shift parameter estimates
obtained independently from VLBA images and single-dish lightcurves
are compared for the 2008 activity period of 3C\,454.3.
Methods to measure the core shift are discussed in
Sect.~\ref{methods}, observational data are described in
Sect.~\ref{obsdata}, Sect.~\ref{coreshiftanalysis} presents the
employed analysis techniques, the results are discussed in
Sect.~\ref{discussion}.
Throughout this work we assume the $\Lambda$CDM cosmology with the
following parameters: $H_0 = 71$\,km\,s$^{-1}$\,Mpc$^{-1}$,
$\Omega_\mathrm{m} = 0.27$, and $\Omega_{\Lambda} = 0.73$
\citep[see][]{2009ApJS..180..330K}, which corresponds to a
luminosity distance of $D_L = 5489$\,Mpc, an angular size distance
of $D_A = 1588$\,Mpc, and a linear scale of 7.7~pc~mas$^{-1}$ at the
source redshift.  We use positively defined spectral index $\alpha=d
\ln S/d \ln \nu$.

\section{Methods to measure the apparent core shift}
\label{methods}

% Ways to measure the core shift
There is a number of ways to observe the core shift and related
opacity-driven effects:
\begin{enumerate}
\item
Direct apparent core position measurement through a
phase-referencing experiment is the most obvious way to measure the
core shift. In this method, the telescopes are switching between the
target source and a nearby reference source (phase calibrator). The
switching time is shorter than the coherence time, hence the
calibrator's phase solution can be applied to the target source.
This allows one to preserve information about the target source
position with respect to the phase calibrator. However, the method
of measuring core shift using phase referencing has a few caveats.
First, the reference source also experiences frequency-dependent
position shift, sometimes even if the core is not the dominating
feature of its structure \citep{2011A&A...535A..24S}. Multiple
reference sources could be used to overcome this problem (P.~Voytsik
et al., in prep.). Second, the technique is not free from
ambiguities arising from modeling the source brightness distribution
with a simplified model (typically, consisting of a few Gaussian
emission components). This is true for the phase-referencing as well
as for all the other VLBI-based core~shift measurement techniques,
because even if the absolute coordinate grid is established through
the use of reference sources, the target source structure still has
to be modeled to determine the position of its core. The
phase-referencing method is also relatively expensive in terms of
observing time.

\item
\label{self-ref} The absolute position information is lost during
the phase self-calibration procedure necessary for high-quality VLBI
imaging. Therefore, it is not known {\it a priori}, how images at
different frequencies should be aligned if phase-referencing is not
applied. However, one may find a reference point in VLBI images that
doesn't change its position with frequency. In the absence of strong
spectral index gradients, optically thin features in the jet may
serve as such reference points. The core position may be measured
with respect to an individual jet feature using a brightness
distribution model \citep{2008A&A...483..759K,2011A&A...532A..38S}
or to a large jet section rich in structure by means of image
cross-correlation
\citep{2008MNRAS.386..619C,2009MNRAS.400...26O,2012MNRAS.420..542A,2012AJ....144..105H,2012A&A...545A.113P}.
Both methods provide comparable results as was shown by
\cite{2012A&A...545A.113P}.

\item
\label{size} The opacity effect also manifests itself as the core
size increase at lower observing frequencies. For a conical jet, its
cross-section size depends linearly on the distance from the cone
apex \citep[e.g.,][]{2011A&A...531A..95F}. In this case, the core
size as a function of frequency should obey a power law with the
same index, $-1/k$, as the core position shift versus frequency
\citep{1979ApJ...232...34B,1981ApJ...243..700K,1994ApJ...432..103U,1998ApJ...494..139J,2008arXiv0811.2926Y}.
If the jet is not conical, the dependencies of core size and core
position shift on frequency will differ (resulting in different
values of $k$). The core size measurements are rarely used to study
opacity because this region is often unresolved by ground-based
VLBI.

\item
\label{delay} Opacity is the reason for time delay of lower
frequency radio emission peaks with respect to the ones at higher
frequencies
\citep[e.g.,][]{1985ApJ...298..114M,1992A&A...254...71V,2011A&A...531A..95F}.
Radio flares are believed to be caused by disturbances traveling
down the jet. The radio flare peak at a given frequency occurs
around the time the disturbance passes the core at this frequency.
Therefore, a time delay between radio lightcurves at different
frequencies may be directly related to the core shift parameters
measured with VLBI \citep{2006A&A...456..105B,2011MNRAS.415.1631K}.
The opacity shift inferred from the time lags can be reconciled with
the core shift measurements from VLBI observations under the
following assumptions. The first one is that a flare at a given
frequency $\nu_\mathrm{obs}$ has its peak quite near the jet
\emph{location}, where we observe the core at this frequency (i.e.\
where the optical depth $\tau_{\nu_\mathrm{obs}}=1$). The second
assumption is that the jet Doppler factor is constant, i.e. the
plasma flow speed and the jet viewing angle are constant, so we can
link measured distance to time. If the flow is accelerated, the
relation between time lags and measured core positions will not be
linear.

\item
A new method allowing one to simultaneously measure the frequency
dependent core position shift and its size change using the radio
intraday variability (IDV) caused by interstellar scintillation was
proposed by \cite{2013ApJ...765..142M}. The position offset between
the scintillating component (presumably, the core) at different
frequencies is causing a time delay between the IDV lightcurves at
these frequencies. This delay is stable on timescales much longer
that the scintillation timescale and thus can be distinguished from
the refractive effects of interstellar medium.  The angular scale of
the scintillating source at a given frequency is estimated from the
IDV variability timescale and parameters of the scattering screen.

\end{enumerate}

\section{Observational data}
\label{obsdata}

% This work

%Frequency= 4.6 GHz  Peak= 3.32 Jy/beam Beam= 3.46x1.52 (mas) at PA= -6.6 deg. First contour= 6.00 mJy/beam, Factor=4.00
%Frequency= 5.0 GHz  Peak= 3.44 Jy/beam Beam= 3.13x1.38 (mas) at PA= -6.4 deg. First contour= 6.00 mJy/beam, Factor=4.00
%Frequency= 8.1 GHz  Peak= 5.87 Jy/beam Beam= 1.78x0.87 (mas) at PA= -0.0 deg. First contour= 6.00 mJy/beam, Factor=4.00
%Frequency= 8.4 GHz  Peak= 6.27 Jy/beam Beam= 1.72x0.82 (mas) at PA= -0.9 deg. First contour= 6.00 mJy/beam, Factor=4.00
%Frequency= 15.4 GHz  Peak= 12.02 Jy/beam Beam= 0.93x0.45 (mas) at PA= -5.9 deg. First contour= 6.00 mJy/beam, Factor=4.00
%Frequency= 23.8 GHz  Peak= 15.77 Jy/beam Beam= 0.67x0.29 (mas) at PA= -11.2 deg. First contour= 6.00 mJy/beam, Factor=4.00
%Frequency= 43.2 GHz  Peak= 23.49 Jy/beam Beam= 0.70x0.58 (mas) at PA= -0.4 deg. First contour= 6.00 mJy/beam, Factor=4.00

\subsection{Single epoch VLBI data}

3C\,454.3 was observed on 2~October 2008 with National Radio
Astronomy Observatory (NRAO) Very Long Baseline Array (VLBA)
simultaneously at seven frequencies ($4.6$, $5.0$, $8.1$, $8.4$,
$15.4$, $23.8$, and $43.2$\,GHz) in the framework of our survey of
parsec-scale radio spectra of twenty $\gamma$-ray bright blazars
\citep{2010arXiv1006.3084S,2010arXiv1001.2591S,2011PhDT.........6S}.
The observation was conducted with 16 on-source scans (each
4-7~minutes long depending on frequency) spread over 9~hours.

The data reduction was conducted in the standard manner using the
\texttt{AIPS} package \citep{1990apaa.conf..125G}. A special
procedure, similar to the one described by
\cite{2011A&A...532A..38S}, was applied to improve amplitude
calibration of the correlated flux density resulting in $\sim 5$\%
calibration accuracy at $4.6$--$15.4$\,GHz range and $\sim 10$\%
accuracy at $23.8$ and $43.2$\,GHz. The \texttt{Difmap} software
\citep{1997ASPC..125...77S} was used for imaging and modeling the
$uv$-data. Details of the employed calibration and analysis
technique were discussed by \cite{2011PhDT.........6S}.

\begin{figure}
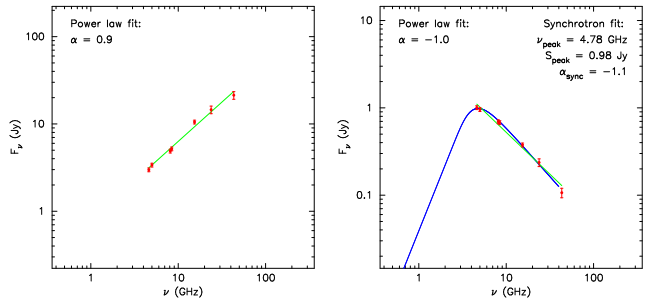

  \centering
  \includegraphics[width=0.22\textwidth, angle=270, trim=0cm 0cm 0cm 0cm, clip]{core_spectrum.ps}
  ~
  \includegraphics[width=0.22\textwidth, angle=270, trim=0cm 0cm 0cm 0cm, clip]{component_spectrum.ps}

 \caption{VLBA spectra of the core (left) and the reference jet component C7 (right) derived from $uv$-modeling.
 The observed spectra are compared to the power law (straight line in log-log scale) and uniform synchrotron cloud models.}
  \label{fig:spectra}
\end{figure}

Fig.~\ref{fig:colormap} shows total intensity VLBA images of
3C\,454.3. While at $4.6$\,GHz the source shows a bright extended
jet, at $43.2$\,GHz its structure is dominated by the bright compact
core. The spectra of the core and the jet component C7 are presented
in Fig.~\ref{fig:spectra}. C7 is the only component that could be
identified across all observing bands (its detection at $43.2$\,GHz
required data tapering). This component has a steep radio spectrum
and, consequently, it is considered to be optically thin. Therefore
C7 may serve as a reference feature for multi-frequency image
alignment. Actually, its spectrum slightly deviates from the power
law at low frequencies and can be fit with the theoretical spectrum
of a uniform synchrotron-emitting cloud. Since C7 is more extended
than the core, its spectrum may be artificially softened due to the
$uv$-coverage related flux density losses at high frequencies. This,
however, does not change the conclusion that C7 is optically thin in
the studied frequency range. The core spectrum is highly inverted
with the spectral slope $\alpha_\mathrm{core}=+0.9$ compatible with
a partially optically thick synchrotron emission of a non-uniform
source.

\subsection{Single-dish radio lightcurves from 4.8 to 37 GHz}

\begin{figure}
  \centering
  \includegraphics[width=0.5\textwidth, angle=0, trim=0cm 0cm 0cm 0cm, clip]{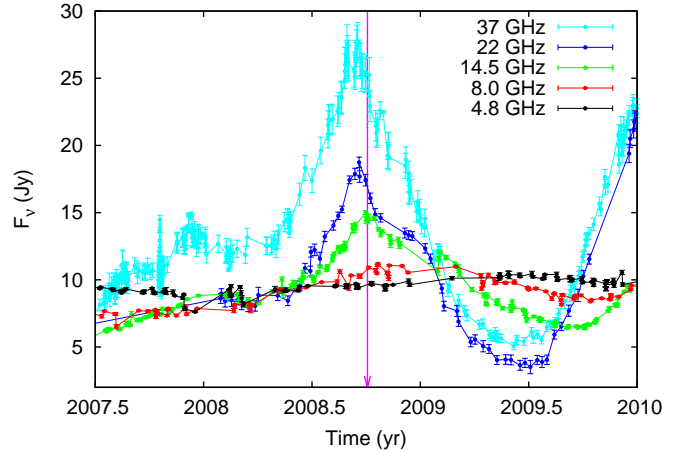}
  \caption{Radio lightcurves of the 2008 flare. The arrow marks the VLBA observation epoch.}
  \label{fig:lc}
\end{figure}

Single-dish flux density monitoring observations of 3C\,454.3 were
obtained with the 26\,m University of Michigan Radio Observatory
(UMRAO) radio telescope at 4.8, 8.0 and 14.5\,GHz, 14\,m
Mets\"{a}hovi telescope at 22 and 37\,GHz and the 22\,m RT-22 radio
telescope of Crimean Astrophysical Observatory (CrAO) at 22 and
37\,GHz. The dataset and the corresponding observing techniques are
presented and described by~\cite{2011ARep...55..608V}. The
lightcurves of the 2008 flare are reproduced by us in
Fig.~\ref{fig:lc}. The flare duration is almost constant across all
bands --- slightly less than two years. One can see that the flare
at lower frequencies is less prominent and delayed in time with
respect to high frequencies with a time lag resulting from the
synchrotron opacity
\citep{1960SvA.....4..243S,1966Natur.211.1131V,1985ApJ...298..114M,2011MNRAS.415.1631K}
as discussed in Sect.~\ref{methods} for the method~\ref{delay}.

\section{Core shift analysis}
\label{coreshiftanalysis}

%\subsection{VLBA analysis}

We characterize the core shift effect using four different
techniques of three methods as described in Sect.~\ref{methods}:
VLBI core position shift measured with brightness distribution
modeling and image cross-correlation (method~\ref{self-ref}), VLBI
core size increase (method~\ref{size}), and time-delay analysis of
single-dish radio lightcurves (method~\ref{delay}).

\subsection{Core position determined from modeling VLBA visibility data}
\label{uv-mod}

\begin{figure}
  \centering
  \includegraphics[width=0.5\textwidth, angle=0, trim=0cm 0cm 0cm 0cm,clip]{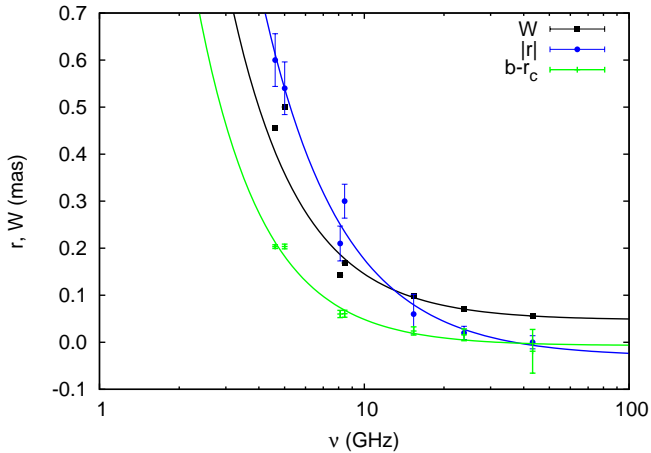}
  \caption{Measured values of the core shift ($|\vec{r}|$), core to C7 component
  separation ($r_\mathrm{c}$), and core width ($W$) as a function of $\nu$
  compared to the corresponding $a \nu^{-1/k}+b$ fits
  (see Table~\ref{table:results}).}
  \label{fig:shifts}
\end{figure}

Following \cite{2008A&A...483..759K} and \cite{2011A&A...532A..38S},
we model jet emission at each frequency with a set of elliptical
Gaussian components in the visibility ($uv$) plane.
\cite{1999ASPC..180..301F} formulas were applied to estimate
position uncertainties of the model components. The well-isolated
component C7 that could be identified across all the observing
frequencies was chosen as the reference for determining the core
position (Fig.~\ref{fig:colormap}).
%A tapering had to be applied to
%visibility data to detect this component at 43\,GHz.
The distance, $\Delta r$, between the apparent core and C7 was
measured at each frequency and fitted by $\Delta
r=\frac{a}{\nu^k}+b$ law where $k=0.56\pm0.22$
(Fig.~\ref{fig:shifts}, the best-fit values of coefficients $a$ and
$b$ are presented in Table~\ref{table:results}). The fact that C7
lies downstream of the region where the jet changes its direction
from West to North-West
%prominent jet band
should not affect the estimated value of $k$. Here and later the
weighted non-linear least-square fitting is performed using the
Levenberg--Marquardt algorithm \citep[e.g,][]{2002nrc..book.....P}
and the reported parameter errors are the asymptotic standard errors
obtained from the variance--covariance matrix after the final
iteration.

\subsection{Core position determined with 2D~cross-correlation image analysis}

3C\,454.3 has a bright and extended parsec-scale jet rich in complex
structures which make it well suited for image alignment with
2D~cross-correlation.  For each pair of adjacent frequencies in our
observation we construct a pair of images convolved with the same
beam (corresponding to the naturally weighted beam at the lower of
the two frequencies).  The pixel size is chosen as 1/20 of the minor
axis of the restoring beam corresponding to the most optimistic
positional accuracy expected for brightest image features. We have
experimented with smaller pixel size but found no improvement in the
image alignment accuracy.

A \texttt{PDL}\footnote{Perl Data Language,
  \url{http://pdl.perl.org/}}-based program written by one of the
authors (TS) was employed to perform the cross-correlation analysis.
The same program was used by \cite{2012AJ....144..105H} and
\cite{2012A&A...545A.113P}. It allows a user to select an image
region that contains complex structures expected to be optically
thin above 5\,GHz. We choose analysis regions containing most of the
visible jet emission at each pair of images.  The optically thick
core region is excluded from the analysis. The process of manual
region selection necessarily introduces a human bias in the process.
In order to minimize this bias, we repeated each measurement
multiple times selecting slightly different analysis regions. The
obtained shifts were verified by visually examining the spectral
index map constructed with the applied shift. The resulting map
should not contain extreme spectral index values, especially near
the edges of emitting regions
\citep{2001ApJ...550..160M,2008A&A...483..759K,2012AJ....144..105H}.
The difference in shifts obtained using different analysis regions,
while the control spectral index map remained acceptable, were no
greater than a few times 1/20 of the minor axis of the
  beam. We adopt this value as an indicator of the cross--correlation
image alignment accuracy for a given frequency pair.

In order to test if the unmatched uv-coverage between the two
frequencies has an impact on our results, for each pair of
frequencies we repeated the analysis fixing the analyzed image areas
and restricting the data to a common uv-range before convolving the
images with the same beam. In all cases, the cross-correlation
results differed by no more than one pixel from the ones obtained
with the images having unmatched uv-coverage and convolved with the
same beam. This confirms that restoring the two images with the same
large beam effectively eliminates the effects of unmatched
uv-coverage on cross-correlation analysis. We also correlated pairs
of images restored with naturally-weighted beams after restricting
the two data sets to the common uv-range. The two beams were not
identical because, while the two data sets had the same uv-range,
the uv-coverage within that range was slightly different. In that
case, the cross-correlation results differed by up to 2 pixels from
the ones obtained with images convolved with identical beams.
Overall, the uncertainty of up to 2 pixels resulting from a choice
of the analysis strategy (identical beams vs. different beams with
matched uv-ranges) is no larger than the one introduced by a choice
of an analyzed image area (a few pixels).

The cross--correlation method allows one to calculate a
displacement, $\Delta\vec{r}_{12}$, between phase centers of images
at two frequencies $\nu_1$ and $\nu_2$. When one knows
$\Delta\vec{r}_{12}$, the core shift is calculated as $\Delta
\vec{r} = \vec{r}_{1} + \Delta\vec{r}_{12} - \vec{r}_2$, where
$\vec{r}_{1}$ and $\vec{r}_2$ are positions of the core relative to
phase center at the two frequencies. Position of the core with
respect to the phase center is measured by modeling the source
structure in \texttt{Difmap} as in Sect.~\ref{uv-mod}.

\cite{2012A&A...545A.113P} reported the core shift of $0.159$\,mas
between 8 and 15\,GHz measured with the VLBA for 3C\,454.3 on
15~June~2006 using the same cross-correlation technique. This is
close to the value obtained in our analysis
(Table~\ref{table:results}, Fig.~\ref{fig:shifts}).

It should be noted that for any given pair of frequencies, the value
of core shift derived from the cross-correlation analysis,
$|\vec{r}|$, is not the same as the difference in core separation
from the jet component C7 derived from visibility model fitting
(Sect.~\ref{uv-mod} and Table~\ref{table:results}, column
$r_\mathrm{c}$). The difference is clearly seen from comparing the
$|\vec{r}|$ and $b$--$r_\mathrm{c}$ measurements and best-fit curves
on Fig.~\ref{fig:shifts}. The reason is that the direction of
$\vec{r}$ differs from the core -- reference C7-component direction.
However, if the angle between the two directions is constant, that
will not affect estimation of the power-law coefficient $k$. The
analysis using the cross-correlation method results in
$k=0.68\pm0.13$.

\begin{table}
 {\centering
 \begin{minipage}{140mm}
 %\begin{minipage}{70mm}
  \caption{Analysis results.}
  \label{table:results}
  %\begin{tabular}{@{}llrrrrlrlr@{}}
  \begin{tabular}{@{}rc@{}c@{}cc@{}}
  \hline\hline
$\nu$~~~ &  $|\vec{r}|$   &    $r_\mathrm{c}$ &   $W$   &  $\Delta T$ \\
  (GHz)  &   (mas)        &      (mas)        &  (mas)  & (year) \\
%  (1)~~ &      (2)       &       (3)         &    (4)  &   (5)  \\
  \hline
  43.2 &  $0.0$          & $5.979 \pm 0.046$ & $0.055$ &  $\dots$ \\
  36.8 & $\dots$         & $\dots$           & $\dots$ &  $0.0$    \\
  23.8 & $0.02 \pm 0.01$ & $5.944 \pm 0.013$ & $0.071$ &  $\dots$ \\
  22.2 & $\dots$         & $\dots$           & $\dots$ &  $0.04 \pm 0.03$   \\
  15.4 & $0.06 \pm 0.04$ & $5.936 \pm 0.009$ & $0.099$ &  $\dots$ \\
  14.5 & $\dots$         & $\dots$           & $\dots$ &  $0.13 \pm 0.03$   \\
   8.4 & $0.30 \pm 0.04$ & $5.899 \pm 0.007$ & $0.169$ &  $\dots$ \\
   8.1 & $0.21 \pm 0.04$ & $5.900 \pm 0.008$ & $0.143$ &  $\dots$ \\
   8.0 & $\dots$         & $\dots$           & $\dots$ &  $0.39 \pm 0.04$   \\
   5.0 & $0.54 \pm 0.06$ & $5.756 \pm 0.005$ & $0.500$ &  $\dots$ \\
   4.8 & $\dots$         & $\dots$           & $\dots$ &  $0.75 \pm 0.07$   \\
   4.6 & $0.60 \pm 0.06$ & $5.757 \pm 0.004$ & $0.456$ &  $\dots$ \\
  %\hline
  \hline
    $a=$  & $6.0\pm2.6$     & $-3.3\pm3.0$      & $4.6\pm2.5$ &     $6.2\pm1.4$ \\
    $b=$  & $-0.03\pm0.03$~~~& $5.96\pm0.04$     & $0.05\pm0.01$ & $-0.06\pm0.02$~~ \\
    $k=$  & $0.68\pm0.13$   & $0.56\pm0.22$     & $0.60\pm0.09$ &  $0.78\pm0.08$ \\
  \hline

\end{tabular}
\end{minipage}
} % centering

{\bf Column designation:} Col.~1~--~frequency, Col.~2~--~core
position shift from cross-correlation analysis, Col.~3~--~core
separation from the reference jet component C7, Col.~4~--~core size
on the half-power level, Col.~5~--~lightcurve time delay with
respect to the 36.8 GHz peak. The last three rows present the best
values for coefficients in the $a \nu^{-1/k}+b$ fit to the data in
the corresponding columns.

\end{table}

\subsection{Core size as a function of frequency}

We fitted the measured major axis of elliptical Gaussian core
components versus frequency dependency with the function $W=a
\nu^{-1/k}+b$ and obtained $k=0.60\pm0.09$
(Table~\ref{table:results}, Fig.~\ref{fig:shifts}), which is
consistent with the values derived from the core position analysis
above. The core size uncertainty estimated following
\cite{1999ASPC..180..301F} is unrealistically small at all
frequencies due to the large core flux density. Therefore we do not
report these error estimates in Table~\ref{table:results}. Clearly,
in this case, the error of the estimated core size is dominated by
modeling uncertainties that are hard to quantify. At each frequency
we check that the estimated core size is larger than the resolution
limit computed following
\cite{2005astro.ph..3225L,2005AJ....130.2473K}.

\subsection{Time lag analysis of single-dish radio lightcurves}

\begin{figure}
  \centering
  \includegraphics[width=0.5\textwidth, angle=0, trim=0cm 0cm 0cm 0cm, clip]{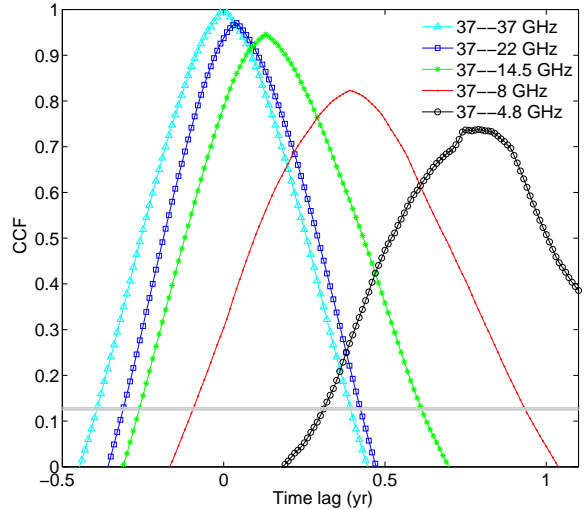}
  \caption{Cross-correlation function for the radio lightcurves.}
  \label{fig:ccf}
\end{figure}

We analyze radio lightcurves of the 2008 flare (Fig.~\ref{fig:lc})
to compare the results with our single-epoch multi-frequency VLBA
observation obtained on 2~October~2008 around the time the flare
peaks at 15\,GHz. Following~\cite{1998PASP..110..660P}, we linearly
interpolate the lightcurves to calculate the corresponding
cross-correlation functions (CCF, Fig.~\ref{fig:ccf}). The
comparison of this method with discrete CCF proposed by
\cite{1988ApJ...333..646E} results in a good agreement. The CCF is
calculated between the lightcurves at 36.8\,GHz and other
frequencies.  The time span of the flare is taken to be
  2~years. For the 36.8\,GHz lightcurve we set it to be
  2007.5--2009.5. Time lags at low frequencies are non-negligible
  compared to the cross-correlation window width. Increasing the
  window width would cause the following 2009 flare, which rises at
  high frequencies, to be included in the analysis, which might affect
  results. Instead, we shift the two-year-wide analysis window for
  each frequency below 36.8\,GHz. We compute CCF values for trial
  shifts in the range from 0 to 1 year. The shift that maximizes the
  CCF value is used to find the time lag between the lightcurves.

To estimate an error of the resulting time lag we used ``FR-RSS''
method described by~\cite{1998PASP..110..660P} consisting of 1000
cycles of Monte~Carlo flux density randomization together with
modified bootstrapping, which allows us to account for estimated
flux density measurement errors, errors due to data sampling and
``outlier'' points. We also add a normally-distributed random time
shifts to the lightcurves within each simulation with a standard
deviation equal to the mean separation between the observations. The
cross-correlation peak distribution was obtained for 37\,GHz with
each frequency and the time lag error was estimated as the standard
deviation of this distribution.

The time lags measured by us for the 2008 activity period (see
Table~\ref{table:results}) are in agreement with the values ($\Delta
T_\mathrm{37-22} = 0.03$\,year and $\Delta T_\mathrm{37-15} =
0.11$\,year) obtained by \cite{2007A&A...464L...5V} for 2005--2007.
See also Fig.~3 in~\cite{2007ARep...51..450V}.
\cite{2007MNRAS.381..797P} report a wide range of time lags measured
in 1990--2001 with typical values close to our results.

\begin{figure}
  \centering
  \includegraphics[width=0.5\textwidth, angle=0, trim=0cm 0cm 0cm 0cm,clip]{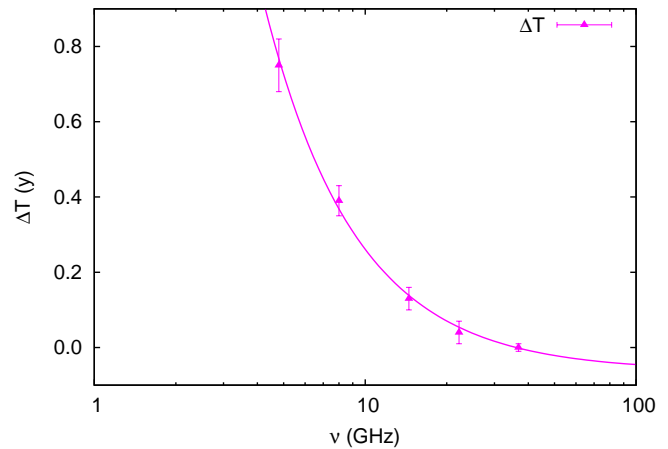}
  \caption{Measured values of the time lag, $\Delta T$, as a function of $\nu$
  compared to the corresponding $a \nu^{-1/k}+b$ fit
  (see Table~\ref{table:results}).}
  \label{fig:timelag}
\end{figure}

Fig.~\ref{fig:timelag} illustrates the frequency dependence of the
time delay with respect to the 37\,GHz lightcurve. We perform a
weighted nonlinear least-squares fit with the model $\Delta T = a
\nu^{-1/k}+b$ which results in $k=0.78\pm0.08$.

\section{Discussion}
\label{discussion}

The \cite{1981ApJ...243..700K} jet model assumes that magnetic field
strength $B$ and particle density $N$ are declining with distance
from the jet origin, $r$, according to the power law: $B =
B_{1}(r/r_{1})^{-m}$, $N = N_{1}(r/r_{1})^{-n}$ (here $B_{1}$ and
$N_{1}$ are values at $r=r_{1}=1$\,pc from the jet base). The power
law index $k$ in the relation of the core position, $r_\mathrm{c}$,
and core size, $W$, change with frequency ($r_\mathrm{c} \propto W
\propto \nu^{-1/k}$) depends on the magnetic field and particle
density indexes $m$ and $n$ and the optically thin spectral index
$\alpha$. If an equipartition fraction or jet power change along the
jet, these changes will also affect the value of $k$
\citep{2013MNRAS.431.1840P}. To estimate $m$ and $n$ we consider two
possible situations: 1)~ambient medium pressure on the jet is
negligible (Sec.~\ref{neglpres}) and 2)~the external pressure is
non-negligible and has a non-zero gradient along the jet
(Sec.~\ref{extpres}). For all calculations in this section we adopt
$k=0.7 \pm 0.1$.

\subsection{Negligible ambient medium pressure}
\label{neglpres}

Assuming that the ambient medium pressure on the jet is negligible
(the ``free jet'' assumption), we have
\begin{equation}\label{eqn_k}
k=[(3-2\alpha)m+2n-2]/(5-2\alpha)
\end{equation}
\citep{1981ApJ...243..700K,1998A&A...330...79L}, where $\alpha$ is
the optically thin spectral index. The assumption of equipartition
between the magnetic field and electron energy densities requires
that
\begin{equation}\label{eqn_equipartition}
N_{1}(r/r_{1})^{-n}\gamma_\mathrm{min} m_e c^2 = K
B_{1}^2(r/r_{1})^{-2 m}/8\pi,
\end{equation}
therefore $n=2m$, where $m_e$ is the electron mass,
$\gamma_\mathrm{min}$ is the minimal Lorentz-factor of emitting
electrons,
$K\approx1/\ln(\gamma_\mathrm{max}/\gamma_\mathrm{min})\approx0.1$
assuming optically thin spectral index $\alpha=-0.5$ %(see the discussion below)
and the maximum Lorentz-factor of emitting electrons
$\gamma_\mathrm{max}=10^{4.34}\gamma_\mathrm{min}$~\citep{2005ApJ...619...73H}.

If the jet flow speed is constant, the jet has a conical shape
(constant opening angle) and the particle density $N \propto r^{-2}$
($n=2$). In that case, the equipartition condition leads to $m=1$
and $k=1$ (for any $\alpha$). However, if jet flow speed and/or its
opening angle are changing along the jet, other combinations of $m$,
$n$, and $\alpha$ are possible that would satisfy the equipartition
condition and result in $k \ne 1$. The values of $m$ and $n$ can be
derived from (\ref{eqn_k}) given the observed value of $k=0.7$ and
assuming $\alpha=-0.5$: $m \simeq 0.8$, $n=2m=1.6$. We note that the
result only weakly depends on the assumed value of $\alpha$.

The core position offset $\Delta r_\mathrm{mas}$ (expressed in
milliarcseconds) between two frequencies $\nu_1$ and $\nu_2$
($\nu_1<\nu_2$, GHz) may be expressed by the parameter
\citep{1998A&A...330...79L}
\begin{equation}\label{eqn_Omega}
    \Omega=4.85\cdot10^{-9}\frac{\Delta
    r_\mathrm{mas}D_\mathrm{L}}{(1+z)^2}\left(\frac{\nu_1^{1/k}\nu_2^{1/k}}{\nu_2^{1/k}-\nu_1^{1/k}}\right),
\end{equation}
where $D_\mathrm{L}$ is the luminosity distance in pc at the
redshift $z$. Averaging over all the frequency pairs and excluding
the close pairs 8.1/8.4 and 4.6/5.0\,GHz we obtain $\Omega = 43 \pm
10$\,pc\,GHz$^{1/k}$.
% ;)
Following \cite{2009MNRAS.400...26O}, we evaluate equation (43) from
\cite{2005ApJ...619...73H}, assuming the equipartition condition
(\ref{eqn_equipartition}), to estimate the magnetic field strength
at 1\,pc from the jet base:
\begin{equation}\label{eqn_B}
B_{1}\approx
0.014\left(\frac{\Omega^{3k}(1+z)^2\ln{\gamma_\mathrm{max}/\gamma_\mathrm{min}}}{\delta^2\phi\sin^{3k-1}\theta}\right)^{1/4},
\end{equation}
where $\theta$~is the jet viewing angle, $\phi$~is the jet half
opening angle and $\delta \equiv [\Gamma (1-\beta \cos
\theta)]^{-1}$~is the Doppler-factor.

\cite{2010ApJ...715..362J} measured a large variety of values of
$\theta$, $\Gamma_\mathrm{j}$, and $\delta$ for three
VLBI-components in 3C\,454.3. They supposed that these components
move along different sides of the jet (closer or farther from the
line of sight). We adopt the average values measured by
\cite{2005AJ....130.1418J}: $\theta=1.3^\circ$, $\phi=0.8^\circ$,
$\delta=24.6$, $\Gamma_\mathrm{j}=15.6$, the value of $k=0.7$ and
$\ln(\gamma_\mathrm{max}/\gamma_\mathrm{min})=10$. Therefore we get
$B_{1}=0.4\pm0.2$\,G. The estimated uncertainty is formally
propagated from the adopted uncertainties of $k$ and $\Omega$. We
note however, that the uncertainties of $\delta$, $\phi$, $\theta$
and $\ln(\gamma_\mathrm{max}/\gamma_\mathrm{min})$ also contribute
to the total uncertainty. The obtained value of $B_{1}$ is close to
the one estimated by \cite{2012A&A...545A.113P} from 8--15\,GHz core
shift measurements with the assumption of $k=1$.
Following~\cite{1998A&A...330...79L} we estimate the core position
at 43\,GHz and, hence, the magnetic field in the 43\,GHz core:
$B_\mathrm{43\,GHz~core}=0.07\pm0.04$\,G. For the 15\,GHz core:
$B_\mathrm{15\,GHz~core}=0.02\pm0.01$\,G.

The de-projected distance of the core at a given frequency from the
central engine may be estimated following
\cite{1998A&A...330...79L,2005ApJ...619...73H}: $r_{\rm core} (\nu)
= \frac{\Omega}{\sin \theta} \nu^{-1/k_{\rm r}}$. For the 43\,GHz
core, assuming the above value of $\theta$, $r_{\rm core} ({\rm
43\,GHz}) \sim 9$\,pc. For the 15\,GHz core, $r_{\rm core} ({\rm
15\,GHz}) \sim 38$\,pc.

We can estimate the apparent jet half-opening angle by
  comparing core sizes and positions measured at different frequencies
  (Col.~4 and 2 in Table~\ref{table:results}) as $\phi_\mathrm{app}
  \approx \Delta W/ 2 \Delta r \approx 17^\circ$, which corresponds to
  the intrinsic half-opening angle $\phi=\phi_\mathrm{app}\sin \theta
  \approx 0.4^\circ$. This is two times smaller than the value used
  above. However our estimate is subject to uncertain modeling errors
  in W and we prefer to use the values of $\phi$ and $\theta$ obtained
  by \cite{2005AJ....130.1418J}. If we use our value of $\phi$
  instead, the magnetic field strength does not change within the
  estimated error.

\subsection{Total jet power}

The total jet power, often referred to as ``kinetic luminosity'',
can be computed within the equipartition assumption using
equation~(46) of \cite{2005ApJ...619...73H}. As input parameters for
the computation we take the above values of $\theta$, $\phi$,
$\delta$, and $\Gamma_\mathrm{j}$ determined by
\cite{2005AJ....130.1418J}, the measured values $k=0.7$, $\alpha =
-0.5$ and assume constant $\Gamma_\mathrm{j}$ along the jet. The two
critical assumptions are the minimum Lorentz factor of the emitting
particles, $\gamma_\mathrm{min} = 100$ \citep{2005ApJ...619...73H},
and jet composition (electron/positron vs. electron/proton plasma).
For an electron/positron jet, the total kinetic power is
a~few~$\times 10^{44}$~ergs~s$^{-1}$. For the same
$\gamma_\mathrm{min}$, the electron/proton jet would be
$m_{p+}/m_{e-}=1836.2$ times more powerful.

%   Omega        Pe        Pp
% 1.5026e+33  8.1586e+44 1.4980e+48
% 9.4667e+32  5.0228e+44 9.2226e+47
% 9.6497e+32  5.1248e+44 9.4099e+47
% 8.2144e+32  4.3275e+44 7.9460e+47

\subsection{External pressure gradient}
\label{extpres}

If the external pressure drops along the jet as $p\propto r^{-a}$,
the jet will be constantly accelerating as $\Gamma_\mathrm{j} =
\Gamma_\mathrm{j\ast}(r/r_\ast)^{a/4}$, where $r$ is axial jet
coordinate and values marked with~$\ast$ are related to the point
where jet becomes supersonic~\citep{1996ASPC..110..262G}. For $r \gg
r_\ast$ the jet cross-section $d \propto r^{a/4}$ (the observed core
size $W \approx d$). The depicted situation is true in a
hydrodynamically accelerating and adiabatic, steady-state jet. In
this case, the core size dependence on the observing frequency
($W(\nu) \propto \nu^{-a/(4k)}$) differs from the one of the core
position offset ($r \propto \nu^{-1/k}$) if $a \neq 4$.

The measured value of $k$ may be related to the pressure gradient
$a$ (see Fig.~1 in \citealt{1998A&A...330...79L}) resulting in $a
\simeq 2.2$. Hence the magnetic field and particle density power-law
indexes become $m_p\approx 0.4$ and $n_p\approx2.2$.  However, the
power law coefficients for $W(\nu)$ and $r(\nu)$ would differ by a
factor of $a/4 \simeq 0.55$ which contradicts the obtained values
(Table~\ref{table:results}). We conclude that the external pressure
gradient is not a dominant factor in determining the jet geometry in
the region of 43--4.6\,GHz core of 3C\,454.3.

\subsection{Jet speed estimated from time lags and core shift}

Assuming that a lightcurve peak at a given frequency, $\nu$, occurs
when a plasma condensation (jet component) traveling down the jet
passes the region of the core at that frequency, $r_\nu$, one can
estimate the apparent projected speed of such a plasma condensation
by comparing the lightcurve time delay, $\Delta T$, with the VLBI
core position shift between a pair of frequencies, $\Delta r$:
$\mu_\mathrm{app}=\Delta r/\Delta T \simeq 0.7$\,mas/yr (the exact
value depends on the choice of the frequency pair). This value is
2--8 times larger than $\mu_\mathrm{app}$ directly measured with
VLBI by \cite{2009AJ....138.1874L} and \cite{2010ApJ...715..362J}.
One possible explanation for this discrepancy is that the 2008 flare
might have been caused by an exceptionally fast jet component. This
seems possible considering the wide distribution of individual
component's velocities observed with VLBI.
On the other hand, the discrepancy may result from the limited
applicability of the above assumption. If the peak flux density of
the flaring component is declining while the flare develops and the
component becomes optically thin at progressively lower frequencies
(as expected for the adiabatic losses dominated stage of flare
development, \citealt{1985ApJ...298..114M}), the lightcurve peak
would occur earlier than the component reaches the position in the
jet marked by the quiescent core at this frequency. This will result
in an overestimated component velocity since the time lag between
peaks will be smaller than the time it takes for the component to
travel between the higher- and lower-frequency core positions.
In principle, one could also suggest that VLBI does not always
measure the apparent speed which represents the true plasma flow
speed. A hint that this might be the case comes from comparing
results of VLBA kinematics measurements in the inner jet of M87 at
2~cm \citep{2007ApJ...668L..27K} and 7~mm
\citep{2007ApJ...660..200L}.

\section{Summary}
\label{summary}

In the jet model of \cite{1979ApJ...232...34B} and
\cite{1981ApJ...243..700K}, synchrotron opacity manifests itself in
the frequency-dependent position and size of the apparent VLBI core,
as well as in the time delay between radio lightcurves obtained at
different frequencies.  We observe these effects in 3C\,454.3 using
4.6--43\,GHz VLBA images and 4.8--37\,GHz lightcurves obtained with
the 26\,m UMRAO, 22\,m CrAO, and 14\,m Mets\"{a}hovi telescopes in
2007--2009.

Our results support this model as an appropriate description of jet
physics in the apparent parsec-scale core region. The distance of
the core from the jet origin $r_\mathrm{c}(\nu)$, the core size
$W(\nu)$, and the lightcurve time lag $\Delta T(\nu)$ all depend on
the observing frequency $\nu$ as $r_\mathrm{c}(\nu) \propto W(\nu)
\propto \Delta T(\nu) \propto \nu^{-1/k}$. We find the value of the
coefficient to be in the range $k=$~0.6--0.8, consistent with the
synchrotron self-absorption being the dominating opacity mechanism
in the jet of 3C\,454.3, as opposed to free-free absorption found in
relativistic jet sources viewed at large angles to the line of
sight, e.g., Cyg~A \citep{2008evn..confE.108B} and NGC\,1052
\citep{2004A&A...426..481K}. Zamaninasab et al. (2013 A\&A
submitted) analyzed two epochs (2005-05-19 and 2009-09-22) of
simultaneous multi-frequency (5--86\,GHz) VLBA observations of
3C\,454.3 and obtained the values of $k=0.9\pm0.2$ and $k=0.8\pm0.3$
for the 2005 and 2009 epochs, respectively. No difference between
the frequency dependence of $r_\mathrm{c}(\nu)$ and $W(\nu)$ is
observed which suggests that the external pressure is not
significant for the jet geometry in the cm-band core region of
3C\,454.3.

Assuming equipartition, we estimate the magnetic field strength
1\,pc from the jet origin to be $B_{1} \sim 0.4$\,G. It scales with
distance from the central engine as $B = 0.4(r/r_{1})^{-0.8}$\,G.
Within the equipartition assumption, the total kinetic power of the
jet, assuming electron/positron composition and $\gamma_\mathrm{min}
= 100$, is a~few~$\times 10^{44}$~ergs~s$^{-1}$. The electron/proton
jet would be about two thousand times more powerful.

The remarkable agreement found between results obtained with
different techniques both improves robustness of presented results
and supports the lightcurve time lag measurements as an efficient
tool to study characteristics of the opaque apparent base of AGN
jets.

\section*{Acknowledgments}

This research has made use of VLBA observations (project code
BK150). The National Radio Astronomy Observatory is a facility of
the National Science Foundation operated under cooperative agreement
by Associated Universities, Inc. This project was partly supported
by the Russian Foundation for Basic Research (projects 12-02-33101
and 13-02-12103), the basic research program ``Active processes in
galactic and extragalactic objects'' of the Physical Sciences
Division of the Russian Academy of Sciences, and the Ministry of
Education and Science of the Russian Federation (agreement
No.~8405). YYK thanks the Dynasty foundation for support. This
research has made use of NASA's Astrophysics Data System. We thank
the referee whose comments helped to improve the paper.

\bibliographystyle{mn2e}
\bibliography{3c454_coreshift}

\label{lastpage}
\end{document}